\def\Jpsi { \mbox{$J \! / \!\!\; \psi$} }
\def\GLG  { \Gamma_{\rm L}/\Gamma }

\documentstyle[aps,prl,preprint,floats,epsf]{revtex}
\input{psfig}

\begin{document}

\preprint{\tighten\vbox{\hbox{\hfil CLNS 96/1455}
                          \hbox{\hfil CLEO 96-24}}}

\title{Measurement of the Decay Amplitudes and Branching Fractions
of $B \to \Jpsi \, K^*$ and $B \to \Jpsi \, K$ Decays}

\author{CLEO Collaboration}

\date{\today}

\maketitle

\tighten

\begin{abstract} 
Using data taken with the CLEO II detector,
we present the first full angular analysis in the color-suppressed
modes $B^0 \to \Jpsi \, K^{*0}$ and $B^+ \to \Jpsi \, K^{*+}$.
This leads to a complete determination of the decay amplitudes of these modes.
In addition, we update the branching fractions for
$B \to \Jpsi \, K$ and $B \to \Jpsi \, K^*$.
\end{abstract}

\newpage
{
\renewcommand{\thefootnote}{\fnsymbol{footnote}}
\begin{center}
C.~P.~Jessop,$^{1}$ K.~Lingel,$^{1}$ H.~Marsiske,$^{1}$
M.~L.~Perl,$^{1}$ S.~F.~Schaffner,$^{1}$ D.~Ugolini,$^{1}$
R.~Wang,$^{1}$ X.~Zhou,$^{1}$
T.~E.~Coan,$^{2}$ V.~Fadeyev,$^{2}$
% ...
\end{center}
 
\small
\begin{center}
$^{1}${Stanford Linear Accelerator Center, Stanford University, Stanford,
California 94309}\\
$^{2}${Southern Methodist University, Dallas, Texas 75275}\\
% ...
\end{center}

\setcounter{footnote}{0}
}
\newpage


One of the interests in $B \to \Jpsi \, K^*$
decays is their role in $ C\!P\,\! $ violation measurements at asymmetric
$B$-factories. 
The vector-vector decay $B^0 \to \Jpsi \, K^{*0}$, with
$ K^{*0} \to K^0_S \pi^0$,
is a mixture of $ C\!P\,\! $-even and $ C\!P\,\! $-odd eigenstates
since it can proceed via an S, P, or D wave decay.
If one $ C\!P\,\! $ eigenstate dominates or if the
two $ C\!P\,\! $ eigenstates can be 
separated, this decay can be used to measure the angle $\beta$
of the unitarity triangle
in a manner similar to which the $ C\!P\,\! $-odd eigenstate
$B^0 \to \Jpsi \, K^0_S$ is used.

Measurements of the decay amplitudes of
$B \to \Jpsi \, K^{(*)}$ transitions also provide a
test of the factorization hypothesis in decays with internal $ W\! $-emission.
Several phenomenological models, based on the 
factorization hypothesis, predict the longitudinal polarization fraction in
$B \to \Jpsi \, K^*$, denoted $\GLG$,
and the ratio of vector to pseudoscalar meson production,
$R \equiv {\cal B}(B \to \Jpsi \, K^*)/{\cal B}(B \to \Jpsi \, K)$
\cite{bsw,isgw,neubert,cddfgn,aleksan}.
It has been noted \cite{aleksan,gourdin}, that form factor models 
cannot simultaneously explain the earlier experimental data
for these two quantities.
The high values of $\GLG$ measured by ARGUS \cite{argus}
and CLEO II \cite{bigb}, with low statistics,
are not consistent with factorization and the measured value of $R$.
The CDF collaboration has measured a lower value of $\GLG$ \cite{cdf}.
Additional information about the validity of factorization
can be obtained by a measurement of the decay amplitude phases,
since any non-trivial phase differences indicate final state interactions
and the breakdown of factorization \cite{koerner}.

In this paper we present a complete angular analysis 
and an update of the branching fractions for
$B \to \Jpsi \, K^{(*)}$ decays
using the full CLEO II data sample.
Assuming isospin symmetry, we determine the fraction of longitudinal
polarization, the parity content and the phase differences 
of the decay amplitudes
from the modes $B^+ \to \Jpsi \, K^{*+}$ and $B^0 \to \Jpsi \, K^{*0}$
using the $K^{*+}$ and $K^{*0}$ decay modes to 
$K^+ \pi^0$, $K^0 \pi^+$, $K^+ \pi^-$, and $K^0 \pi^0$.
The $\Jpsi$ is reconstructed in its leptonic decay modes
to $e^+ e^-$ and $\mu^+ \mu^-$.
The measurements presented here supersede previous CLEO II results
\cite{bigb}, which are based on a subset of the data used for this analysis.
  
The decay $B \to \Jpsi \, K^*$ is described by three complex decay amplitudes. 
Following a suggestion of Dunietz {\it et al.} \cite{dunietz,dighe},
we measure the decay amplitudes 
$A_0 = - \sqrt{1/3} \, S + \sqrt{2/3} \, D$, 
$A_\parallel = \sqrt{2/3} \, S + \sqrt{1/3} \, D$, and $A_\perp = P$,
where $S$, $P$, and $D$ denote S, P, and D wave amplitudes, respectively.
Normalizing the decay amplitudes to
$|A_0|^2 + |A_\parallel|^2 + |A_\perp|^2 = 1$ 
and eliminating one overall phase leaves four independent parameters.

The full angular distribution of a $B$ meson decaying into two vector particles
is specified by three angles.
Previously the helicity angle basis \cite{wick} has been used for 
angular analyses of $B \to \Jpsi \, K^*$ decays. 
Because of its convenience for extracting the parity information,
we use a different set of angles, called the transversity basis \cite{dighe}.
The direction of the $K^*$ in the $\Jpsi$ rest frame defines
the x-axis of a right-handed coordinate system.
The $ K \pi$ plane fixes the y-axis with $p_y(K) > 0$
and the normal to this plane defines the z-axis.
The transversity angles $\theta_{\tiny{\rm tr}}$ and $\phi_{\tiny{\rm tr}}$
are then defined as polar and azimuth angles of the $l^+$ in the
$\Jpsi$ rest frame.
The third angle, the $K^*$ decay angle $\theta_{K^*}$, is defined as that
of the $K$ in the $K^*$ rest frame relative 
to the negative of the $\Jpsi$ direction in that frame.
Using these definitions the full angular distribution of the 
$B \to \Jpsi \, K^*$ decay is \cite{dighe}:

\newpage                     

\begin{eqnarray*}
& & \frac{1}{\Gamma} \:
\frac{{\rm d}^3 \Gamma}{{\rm d} \cos \theta_{\tiny{\rm tr}} \:
                        {\rm d} \cos \theta_{K^*} \: 
                        {\rm d} \phi_{\tiny{\rm tr}} } \nonumber \\*[1mm] 
& & \quad = \; {\textstyle \frac{9}{32 \pi}} \: \{ 
       2 \; |A_0|^2 \; \cos^2 \theta_{K^*} ( 1 - \sin^2 \theta_{\tiny{\rm tr}}
                  \cos^2 \phi_{\tiny{\rm tr}} ) \nonumber \\
& & \quad \quad \quad \quad \;
       + \; |A_\parallel|^2 \; \sin^2 \theta_{K^*} ( 1 - \sin^2 \theta_{\tiny{\rm tr}} 
         \sin^2 \phi_{\tiny{\rm tr}} ) \nonumber \\
& & \quad \quad \quad \quad \;
          + \; |A_\perp|^2 \; \sin^2 \theta_{K^*} \sin^2 \theta_{\tiny{\rm tr}}
              \sin^2 \phi_{\tiny{\rm tr}}  \nonumber \\
& & \quad \quad \quad \quad \;
          - \; {\rm Im} \, (A_\parallel^* A_\perp)
             \; \sin^2 \theta_{K^*} \sin 2 \theta_{\tiny{\rm tr}}
                  \sin \phi_{\tiny{\rm tr}}     \nonumber \\
& & \quad \quad \quad \quad \;
          + \, {\textstyle \frac{1}{\sqrt{2}}} \; {\rm Re} \, (A_0^* A_\parallel)
                \; \sin 2 \theta_{K^*} \sin^2 \theta_{\tiny{\rm tr}}
                  \sin 2 \phi_{\tiny{\rm tr}}    \nonumber \\
& & \quad \quad \quad \quad \;
         + \, {\textstyle \frac{1}{\sqrt{2}}} \; {\rm Im} \, (A_0^* A_\perp)
                \; \sin 2 \theta_{K^*} \sin 2 \theta_{\tiny{\rm tr}}
                                          \cos \phi_{\tiny{\rm tr}} \; \}.
\end{eqnarray*}
For $\bar{B}$ decays the interference terms containing $A_\perp$ switch sign while
all other terms remain unchanged.  

The data for this analysis were recorded with the CLEO II detector
located at the Cornell Electron Storage Ring (CESR). 
We have used a data sample of approximately $3.4 \times 10^6 \ 
B \bar B$ events taken on the $\Upsilon(4S)$ resonance and
representing an integrated luminosity of $3.1~ \rm fb^{-1}$. 
To evaluate non-$b\bar{b}$ backgrounds,
we have also collected a ``continuum'' data sample
60 MeV below the $\Upsilon{\rm (4S)}$\ resonance,
with an integrated luminosity of about $1.6~ \rm fb^{-1}$.

The components of the CLEO II detector \cite{detector} most relevant to this
analysis are the charged particle tracking, the CsI electromagnetic calorimeter
and the muon counters. The tracking system comprises a set of 
precision drift chambers totaling 67 layers inside a 1.5 T solenoidal magnet.
It measures both momentum and specific ionization ($dE/dx$) of charged particles.

Electron candidates are identified by their energy
deposition in the calorimeter, which must equal their measured momenta,
and their specific ionization, which
must be consistent with that expected for electrons.
At least one muon candidate is required to have penetrated five
nuclear interaction lengths of material while the other must have
penetrated at least three interaction lengths.
The decays $B^+ \to \Jpsi \, K^+$
and $B^0 \to \Jpsi \, K^0_S$ have 
little background, therefore only one of the two leptons has to be positively
identified.
In this case the identified muon candidate must penetrate at least
three interaction lengths.
We require the dimuon invariant mass to be within 45 MeV$/c^2$ of the
$\Jpsi$ mass, which corresponds to a $3\sigma$ selection.
For the dielectron invariant mass we require
$-150 \: {\rm MeV}/c^2 < m_{ee} - m_{J\!/\!\!\;\psi} < 45 \: {\rm MeV}/c^2$
to allow for the radiative tail.
The $\Jpsi$ energy resolution is improved by a factor of $5 - 6$
by performing a kinematic fit of the dilepton mass to the nominal $\Jpsi$ mass.
The kinematic fit does not affect the measurements of the transversity angles.
The resolution of the
angle measurements is better than 0.06 radian for all decay angles.

We require the charged hadron candidates to have $dE/dx$ measurements that lie
within three standard deviations ($\sigma$) of the expected values. We
reconstruct $K^0_S$ candidates through the decay to $\pi^+\pi^-$ and
we reconstruct $\pi^0$ candidates through the decay to $\gamma\gamma$.  
Candidate $K^*$ mesons are required to have a $K \pi$ invariant mass 
within 75 MeV$/c^2$ of the nominal $K^*$ mass. 

In symmetric $e^+e^-$ annihilations at the $\Upsilon$(4S) resonance,
the energy of a $B$ meson
must equal the beam energy. We require the energy difference $|\Delta E|$
between the $B$ candidate and the beam energy to be less than
45 MeV for $\Jpsi \, K^+$ and $\Jpsi \, K^0_S$, less than
30 MeV for $\Jpsi \, (K^+ \pi^-)$ and $\Jpsi \, (K^0_S \pi^+)$, and less than
60 MeV for $\Jpsi \, (K^+ \pi^0)$ and $\Jpsi \, (K^0_S \pi^0)$.
These ranges correspond to approximately $3\sigma$ in $|\Delta E|$.
In the rare case, that an event has more than one candidate per mode
we keep only the candidate with the highest probability,
which is based on the measured $|\Delta E|$ and, if available, 
the measured $dE/dx$, the $\pi^0$ mass, and the time-of-flight information.
The resolution on the beam energy is an order of magnitude
better than the resolution
on the $B$ candidate energy. Therefore we substitute the beam energy in the
calculation of the $B$-candidate mass (referred to as the
``beam-constrained mass'' $m_B$).
The detection efficiencies range from 48\% for the
$B^+ \to \Jpsi \, K^+$ mode
down to 9\% for $B^0 \to \Jpsi \, K^{*0}$
with $K^{*0} \to K^0_S \pi^0$.

The most severe background in the $B \to \Jpsi \, K^*$ modes
is feed-across from one $B \to \Jpsi \, K^*$ mode to another.
For such events both the total
energy and the beam-constrained mass are very close to the signal region.
The biggest source of feed-across background is from swapping a random
or misidentified slow $\pi^0$ for the correct one. Consequently most background
events have the $\pi^0$ moving backwards with respect to the $K^*$ 
direction of flight.
To suppress this background we require the $K^*$ decay angle 
to satisfy $\cos{\theta_{K^*}} < 0.7$ in these decays.
This is equivalent to a constraint on the $\pi^0$ momentum, corresponding to a
minimum $p_{\pi^0}$ of about 200 MeV$/c$.
The total fraction of feed-across events in the signal region,
averaged over all $K^*$ modes, is $8.0\%$.

\begin{figure}[thp]
\centerline{\psfig{file=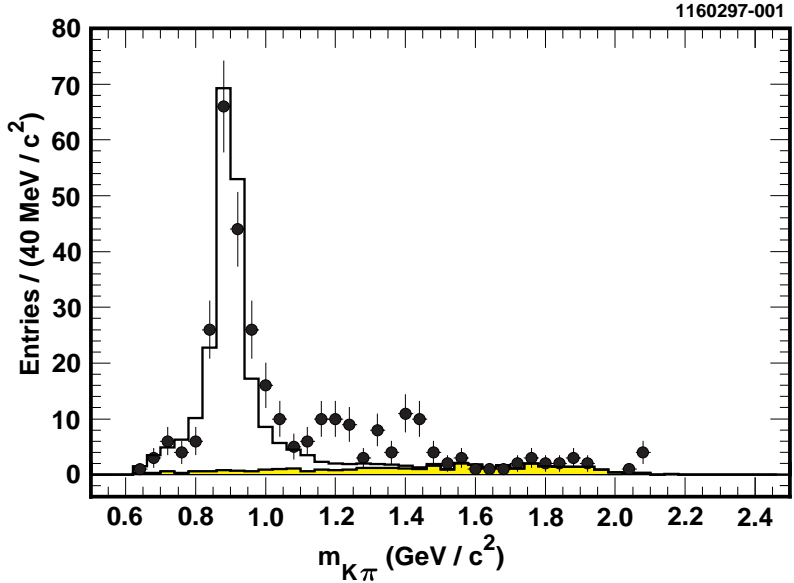,width=0.65\textwidth}}   
\caption{The $m_{K \pi}$ distribution for $m_B > 5.27 \: {\rm GeV}/c^2$.
The data points, the fitted $K^*(892)$ mass peak including
feed-across (empty histogram),
and the combinatorial background (shaded histogram) are shown.}
\label{fig:kpi_mass}
\end{figure}

There might be a contribution from non-resonant $B \to \Jpsi \, K \pi$
decays in the $K^*$ signal region, though neither the previous CLEO
measurement \cite{bigb} nor CDF \cite{cdf} found events in the
$K^*$ mass sidebands.
However, examining the $K \pi$ invariant mass spectrum
(Fig.\ \ref{fig:kpi_mass}) shows an excess of events
between 1.1 and 1.45 GeV$/c^2$.
By computing the kinematics of non-resonant
$B \to \Jpsi \, X_s$ decays, using both the $\Jpsi$ momentum spectrum
from inclusive $B$ decays \cite{psi_incl} 
and several theoretical models \cite{palmer}, we do expect
strangeness-containing final states with invariant masses in this region.
Decays via higher $K^*$ resonances may have line shapes consistent with
the $m_{K\pi}$ distribution seen by us \cite{lass}.
Unfortunately, due to the limited statistics for
$m_{K\pi} > 1.1 \: {\rm GeV}/c^2$,
we cannot distinguish between possible components. 
In addition, we considered many other possible origins for these events
including feed-down from modes such as
$B \to \Jpsi \, K^* \pi$ or $B \to \Jpsi \, K \rho$, and feed-through from
$B \to \Jpsi \, K$, and found none of these
to contribute significantly. 
We estimate the amount of the non-$K^*(892)$ contribution
in the signal region to be $6.4\%$ with a conservatively chosen systematic
uncertainty of $\pm 100\%$.

Other misidentified $B \to \Jpsi \, X$ decays,
like $B \to \Jpsi \, K^* \pi$,
do not contribute significantly to the background since they lie
outside the energy window. With a similar analysis CLEO has found 9 events
for $B^0 \to \Jpsi \, \rho^0$\cite{PsiRho}.
If a pion from the $\rho^0$ is
misidentified as a kaon, $m_{K \pi}$ could fall in the $K^*$ region but
these events would fail the $|\Delta E|$ energy criterion. Feed-across
between the $\Jpsi \, K$ and $\Jpsi \, K^*$ modes is also suppressed by
the requirement on $|\Delta E|$.  
Furthermore, the contributions are uniform in the beam-constrained mass.

We define combinatorial backgrounds to be events that do not contain
a true $\Jpsi \to l^+ l^-$ decay.
In both the $B\bar{B}$ Monte Carlo simulation
and our continuum data sample we see very few such events.

We must correct our data for detection efficiency.
To obtain the efficiency as a function of all three angles,
a large Monte Carlo
sample (120,000 events/$K^*$ mode) is divided into a $20 \! \times \! 20 \!
\times \! 10$ grid in $\cos \theta_{\tiny{\rm tr}}$, $\cos \theta_{K^*}$
and $\phi_{\tiny{\rm tr}}$.
For each $\Jpsi \, K^*$ final state the efficiency is fitted
separately with polynomials in three dimensions including all correlations.  
The efficiency distributions are nearly uniform in all angles except the 
$K^*$ decay angle, where it drops at high 
$\cos \theta_{K^*}$ because of the slow pion.

\begin{figure}[thp]
\centerline{\psfig{file=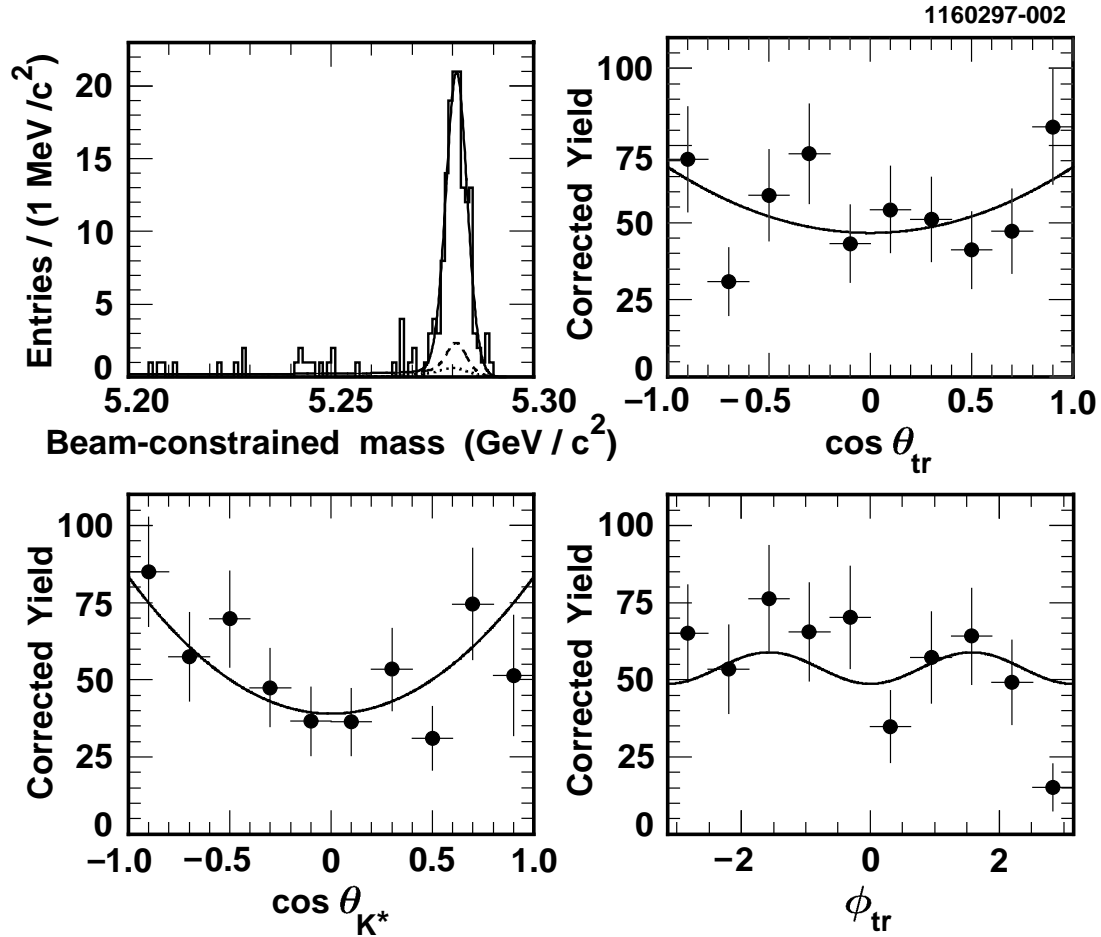,width=0.9\textwidth}}   
\caption{One-dimensional projections of the four-dimensional fit 
to the $B \to \Jpsi \, K^*$ data.
The plot of the beam-constrained mass shows the data (histogram), 
the fit (solid line), the sum of all backgrounds (dashed) and the 
$B \to \Jpsi \, K^*$ feed-across (dotted).
The angular distributions are background subtracted and efficiency corrected.}
\label{fig:result_fit}
\end{figure}

To determine the decay amplitudes, 
a four-dimensional unbinned maximum likelihood fit is performed
to the distributions of the three angles and the beam-constrained mass.
Setting $\phi(A_0) \equiv 0$, we fit for
the longitudinal polarization fraction, 
$|A_0|^2 = \GLG$, the parity-odd fraction,
$|A_\perp|^2 = |P|^2$, and the phases $\phi(A_\parallel)$ and $\phi(A_\perp)$.
Other free parameters in the fit are the branching fraction 
${\cal B}(B \to \Jpsi \, K^*)$, the mean of the $m_B$ distribution
and the normalization of the combinatorial background of each mode.
Fitting for the branching fraction and the polarization parameters
simultaneously ensures the correct treatment of the background events and
automatically adjusts the branching fraction measurement 
for the polarization dependence of the efficiency.
The one-dimensional projections of the resulting fit function are shown in Fig.\ 
\ref{fig:result_fit}.
The results are listed in Table \ref{tab:result_fit}.
The correlations between the fit parameters are small.
The systematic uncertainties of the decay amplitude measurements are dominated 
by those in the efficiency parameterization and background polarization
and are small compared to the statistical errors.

\renewcommand{\arraystretch}{1.2}
\begin{table}[htp]
\begin{center}
\begin{tabular}{cc}
Parameter  & Value \\ \hline
$|A_0|^2 \; = \;
       \Gamma_{\rm L} \: / \: \Gamma$ & $ 0.52 \: \pm \: 0.07 \pm \: 0.04 \quad $ \\
$|A_\perp|^2 \; = \; |P|^2$           & $ 0.16 \: \pm \: 0.08 \pm \: 0.04 \quad $ \\
$\phi(A_\perp)$        & $\,   -0.11 \: \pm \: 0.46 \pm \: 0.03 \; $ radian \\
$\phi(A_\parallel)$    & $\quad 3.00 \: \pm \: 0.37 \pm \: 0.04 \; $ radian \\
\end{tabular}
\end{center}
\caption{Resulting decay amplitudes from the fit to the
transversity angles. The phase $\phi(A_0)$ has been set to zero.
The first error is statistical and
the second is the estimated systematic uncertainty.}
\label{tab:result_fit}
\end{table}
\renewcommand{\arraystretch}{1.0}

We also repeated the fit to the decay amplitudes
using helicity angles rather than transversity angles as well as
performing one-dimensional fits to both the 
longitudinal polarization fraction and the parity-odd component.
An independent angular analysis with the same data sample has also been performed,
using a Monte Carlo technique \cite{stevefit} to evaluate the likelihood 
function. 
All results are in agreement with those reported here. 

These results are the first determination of the parity-odd component 
and the phases of the decay amplitudes of the
$B \to \Jpsi \, K^*$ decay.
The small fraction of the parity-odd component encourages using
the $B^0 \to \Jpsi \, K^0_S \pi^0$ decay
for $ C\!P\,\! $ violation studies at asymmetric $B$ factories.
The phases of the decay amplitudes are measured to be close to zero or $\pi$,
giving no evidence for strong final state interactions. 

\begin{figure}[tbh]
\centerline{\psfig{file=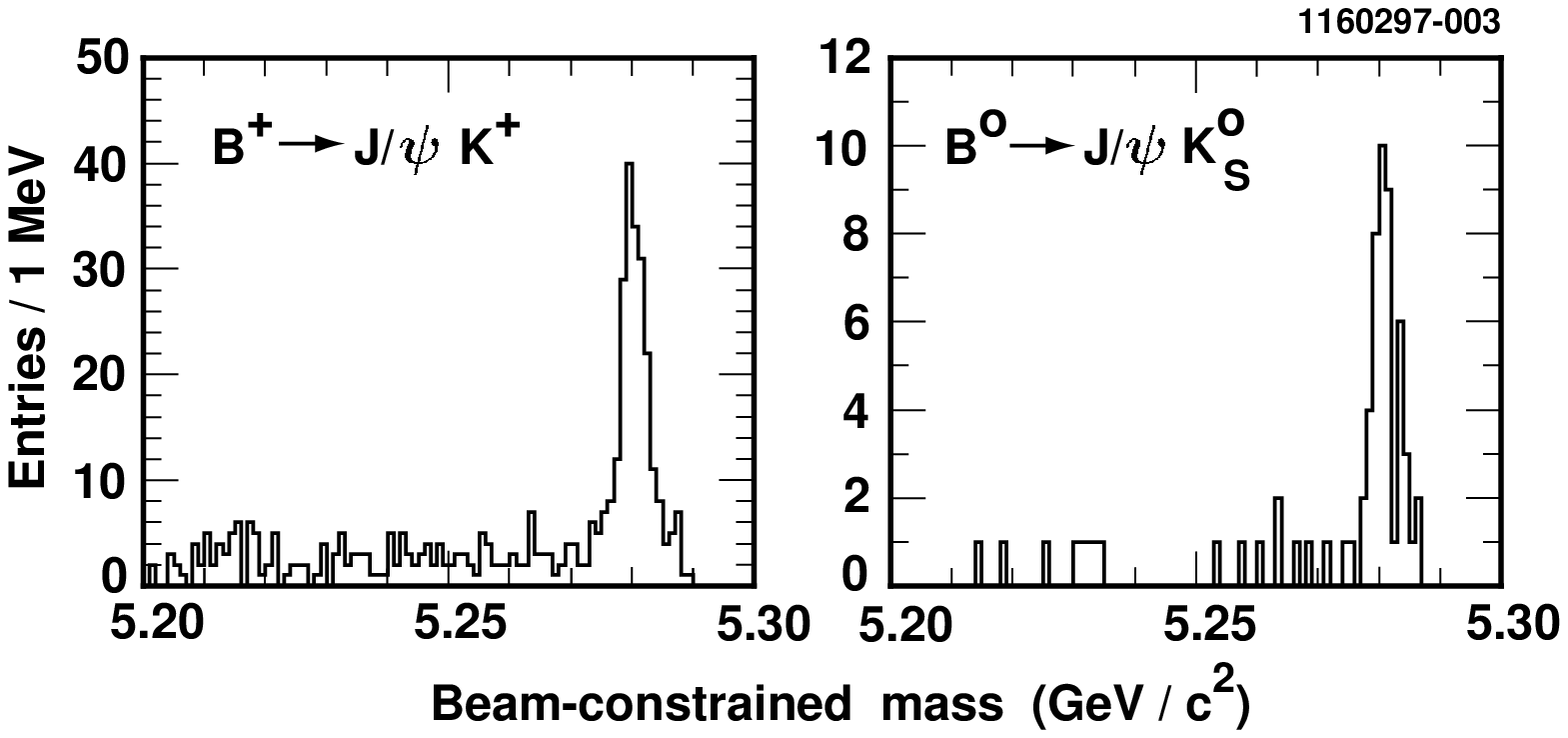,width=0.9\textwidth}}  
\caption{Beam-constrained mass distri\-bu\-tions for the decays 
$B^+ \to \Jpsi \, K^+$ and $B^0 \to \Jpsi \, K^0$.}
\label{fig:result_psi_k}
\end{figure}

The branching fractions of the $B \to \Jpsi \, K^*$ decays
are a result of the angular fit.
To measure the $B \to \Jpsi \, K^+/K^0$ branching fractions
we performed one-dimensional fits to the beam-constrained mass distributions.  
The results of the fits are shown in Fig.\ \ref{fig:result_psi_k}.
All measured branching fractions are listed in Table \ref{tab:result_br}, where
we have assumed that the production rate of neutral and charged $B$
mesons is the same on the $\Upsilon(4S)$ resonance,
in agreement with the actual measured value of 
$f_{+-}/f_{00} = 1.12 \pm 0.20$ \cite{fplusf0}
and a theoretical prediction \cite{lepage}.
The main sources of systematic uncertainties of the $B \to \Jpsi \, K$
branching fraction measurements are track finding, track fitting, 
and lepton identification efficiencies, and the uncertainty 
of the world average of
${\cal B}(\Jpsi \to l^+ l^-)$ \cite{pdg}.
In the $B \to \Jpsi \, K^*$ branching ratios,
uncertainties in the amount of feed-across and non-$K^*$ decays
dominate the systematic error.

\renewcommand{\arraystretch}{1.1}
\begin{table}[tbh]
\begin{center}
\begin{tabular}{ccc}
Decay mode & Signal Yield & Branching Fraction [$10^{-3}$] \\ \hline 
$B^+ \to \Jpsi \, K^+$    & $198.1 \: \pm \: 14.9$
                          & $ 1.02 \: \pm \: 0.08 \: \pm \: 0.07 $ \\
$B^0 \to \Jpsi \, K^0$    & $45.5 \: {+7.3 \atop -6.6} $
                          & $ 0.85 \: {+0.14 \atop -0.12} \: \pm \: 0.06 $ \\
$B^+ \to \Jpsi \, K^{*+}$ & $42.5 \: \pm \: 7.1$
                          & $ 1.41 \: \pm \: 0.23 \: \pm \: 0.24 $ \\
$B^0 \to \Jpsi \, K^{*0}$ & $81.6 \: \pm \: 10.3$
                          & $ 1.32 \: \pm \: 0.17 \: \pm \: 0.17 $ \\
\end{tabular}
\end{center}
\caption{Measured signal yields and branching fractions.}
\label{tab:result_br}
\end{table}
\renewcommand{\arraystretch}{1.0}

With the assumption of equal partial widths, 
$\Gamma(B^+ \to \Jpsi \, K^{(*)+}) = \Gamma(B^0 \to \Jpsi \, K^{(*)0})$,
and eliminating common systematic uncertainties we determine
\begin{displaymath}
\frac{f_{+-}}{f_{00}} \frac{\tau_{B^+}}{\tau_{B^0}}
          \; = \; 1.15 \: \pm \: 0.17 \: \pm \: 0.06 \, .
\end{displaymath}

Assuming isospin invariance, we find
for the ratio of pseudoscalar to vector meson production
\begin{displaymath}
R \; = \; \frac{{\cal B}(B \to \Jpsi \, K^*)}{{\cal B}(B \to \Jpsi \, K)}
           \; = \; 1.45 \: \pm \: 0.20 \: \pm \: 0.17 \, .
\end{displaymath}

These new measurements of $\GLG$ and the ratio $R$ indicate
that the discrepancy with naive factorization models is not as acute as before.

We gratefully acknowledge the effort of the CESR staff in providing us with
excellent luminosity and running conditions.
This work was supported by 
the National Science Foundation,
the U.S. Department of Energy,
the Heisenberg Foundation,  
the Alexander von Humboldt--Stiftung,
%
Research Corporation,
the Natural Sciences and Engineering Research Council of Canada,
and the A.P. Sloan Foundation.


\end{document}